\documentclass[twocolumn,twoside]{revtex4}
\usepackage{graphicx}
\usepackage{fancyhdr}
\begin{document}

\title{Charm lifetime measurements at Belle II}

%

\author{J.~V.~Bennett}
\affiliation{University of Mississippi, University, MS, 38677}

\author{\textbf{on behalf of the Belle~II collaboration}}

\begin{abstract}
Upgrades at the Belle II experiment and the SuperKEKB asymmetric-energy electron-positron collider
enable precise measurements of particle decays. Even with early data, Belle~II has made several
world-leading measurements of particle lifetimes, which are useful as tests of effective models used in
searches for physics beyond the standard model. Especially for charm hadrons, these effective models
depend on careful consideration of non-perturbative effects to give an adequate description of lifetimes.
Recent measurements at other experiments have stimulated particular interest in charm baryons, 
including several measurements that indicate the hierarchy of lifetimes is different than once believed.
The measurements of the $D^+$, $D^0$, and $\Lambda_c^+$ lifetimes at Belle~II are consistent with 
previous measurements, but with improved precision. These measurements will improve the world-average
values, provide refined tests for effective models, and serve as benchmarks for future measurements.

\end{abstract}

\maketitle

\thispagestyle{fancy}


\section{Introduction}
Searches for physics beyond the standard model (SM) of particle physics are complicated by the
need for accurate theoretical descriptions of strong interactions at low energy. This is typically
achieved with predictive tools like the heavy quark expansion 
(HQE)~\cite{Neubert,Uraltsev,Lenz,Lenz15,Kirk,Cheng18,Gratex}. In this framework, decay widths of
hadrons containing a heavy quark, $Q$, are calculated with an expansion in terms of the heavy
quark mass, $m_{Q}$. Particle lifetimes, which are the inverse of the particle decay widths, are
sensitive to higher-order contributions~\cite{Lenz,Cheng15} and therefore provide excellent tests
for effective models. Charm hadrons, in particular, require careful consideration of model-dependent
effects. Non-perturbative effects are relatively small for bottom hadrons, but are significant for charm
hadrons, for which the HQE to 1/$m_{c}^3$, where $m_{c}$ is the mass of the charm quark, does not
give a good description of lifetimes~\cite{Cheng18,Gratex}. Precise measurements of charm hadron
lifetimes therefore provide stringent tests of effective models.

The world average values for charm hadron lifetimes are dominated by measurements that were made 
two decades ago. The $D^0$ and $D^+$ lifetimes were measured at the per-mille level by the 
photon-beam experiment FOCUS~\cite{PDG,Link}, the hyperon-beam experiment SELEX~\cite{Kush}, and the 
electron-positron ($e^+e^-$) experiment CLEO~\cite{Bonv}. These experiments also made percent-level 
measurements of the lifetime of the $\Lambda_c^+$~\cite{Bonv,Link02,Mahm}. More recently, the LHCb 
collaboration made precise measurements of the lifetimes of several charm baryons, relative to the $D^+$ 
lifetime~\cite{Aaij}. Indeed, these relative measurements, which benefit from reduced uncertainties related 
to event selection that may bias the decay time at hadron colliders, are systematics dominated due to
the uncertainty on the lifetime of the $D^+$. Additionally, the recent measurements by LHCb result in
a hierarchy of charm baryon lifetimes that is in conflict with that from previous measurements~\cite{Aaij18,Aaij22}.
Precise, absolute measurements of charm hadrons by the Belle~II experiment help to clarify this
picture and substantially improve the world average values.

\section{The Belle~II Experiment}
Experiments at $e^+e^-$ colliders have an excellent potential for absolute lifetime measurements due
to the ability to reconstruct high charm-hadron yields without biasing the measured decay time. The
Belle~II experiment at the SuperKEKB asymmetric energy $e^+e^-$ collider~\cite{BelleII,SuperKEKB}
benefits from excellent vertexing capabilities that provide a better decay-time resolution than the
previous $e^+e^-$ experiments at Belle and BaBar. The Belle~II detector~\cite{BelleII} includes an upgraded
tracking system comprised of a two-layer silicon pixel detector, a four-layer double-sided silicon strip
detector, and a 56-layer central drift chamber. This tracking system enables very good vertex resolution,
which when coupled with a small beam size and excellent alignment and calibration, enable precise,
absolute measurements of particle lifetimes. Lifetimes are calculated from the distance between the
production and decay vertices. In effect, lifetime measurements are therefore an excellent probe of
the beam spot and alignment calibrations using early Belle~II data.

The SuperKEKB collider provides world-leading instantaneous luminosity according to the nano-beam
scheme, in which the size of the electron and positron beams in the plane transverse to their
momentum is reduced by an order of magnitude compared to those at the KEKB collider~\cite{SuperKEKB}. 
This smaller beam size better constrains the event kinematics and allows for an improved decay-time
resolution for particles. The interaction region is defined by the overlap of the beams, called the beam spot,
which is calibrated continuously using $e^+e^-\rightarrow\mu^+\mu^-$ events.

\section{Lifetime measurements}
\begin{figure}
\includegraphics[width=\linewidth]{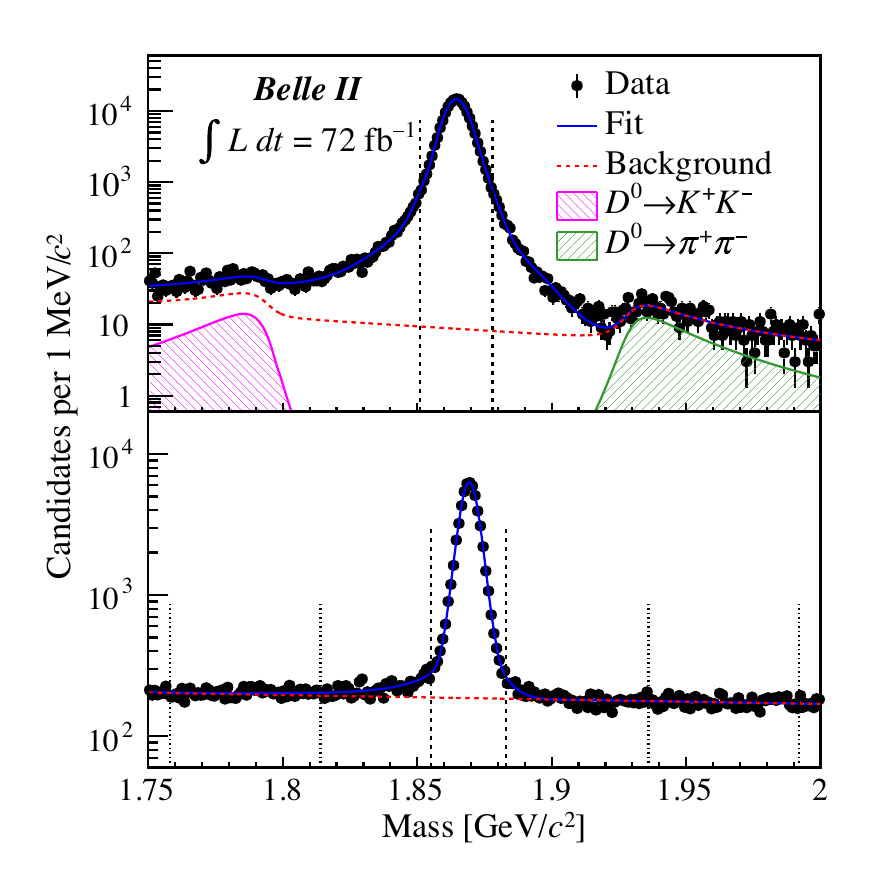}
\caption{\label{dmass} Invariant mass distributions for $D^0\rightarrow K^-\pi^+$ (top) and 
$D^+\rightarrow K^-\pi^+\pi^+$ (bottom) candidates with fit projections overlaid~\cite{D}. The vertical dashed
lines enclose the signal region. The vertical dotted lines in the bottom figure enclose the sideband regions.}
\end{figure}

\begin{figure}
\includegraphics[width=0.9\linewidth]{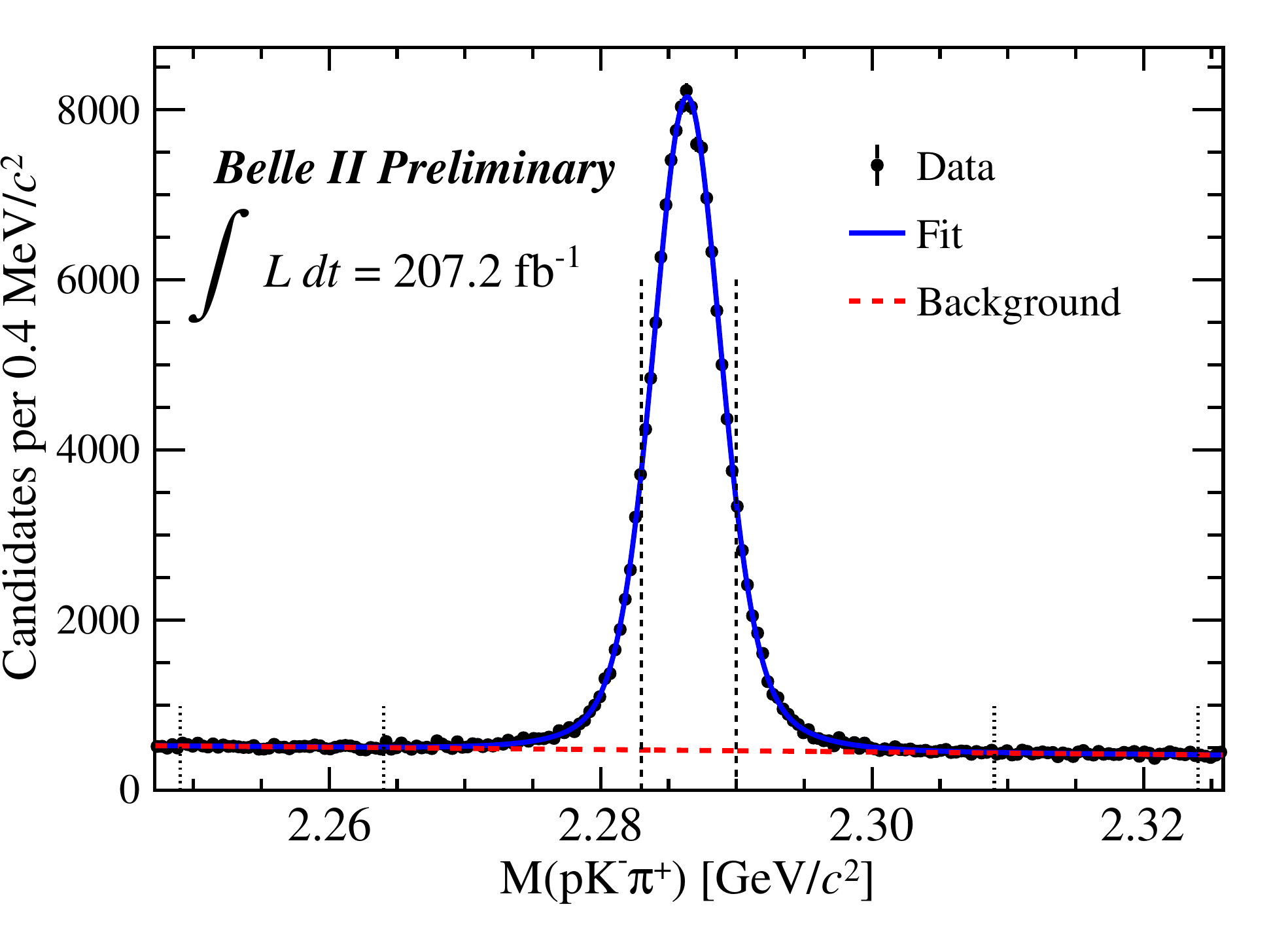}
\caption{\label{lcmass} Invariant mass distributions for $\Lambda_c^+\rightarrow pK^-\pi^+$ with fit 
projections overlaid~\cite{Lc}. The vertical dashed lines enclose the signal region and the dotted lines enclose the 
sideband regions.}
\end{figure}

Precision measurements of the lifetimes of the $D^0$, $D^+$, and $\Lambda_c^+$ are made at Belle~II
using unbinned, two-dimensional, maximum-likelihood fits to the distributions of the decay-time t and
decay-time uncertainty $\sigma_t$~\cite{D,Lc}. Backgrounds are neglected for the $D^0$ lifetime measurement, but
the $D^+$ and $\Lambda_c^+$ lifetime measurements account for backgrounds that remain after
event selection by applying a simultaneous fit to the events in the signal and sideband invariant-mass 
regions. Fits to the mass distributions, as shown in Fig.~\ref{dmass} and Fig.~\ref{lcmass} for the $D^{0,+}$ and $\Lambda_{c}^{+}$
measurements, respectively, are used to determine the rate of background contamination, which is applied as a constraint in the lifetime fit.

\begin{figure}
\includegraphics[width=\linewidth]{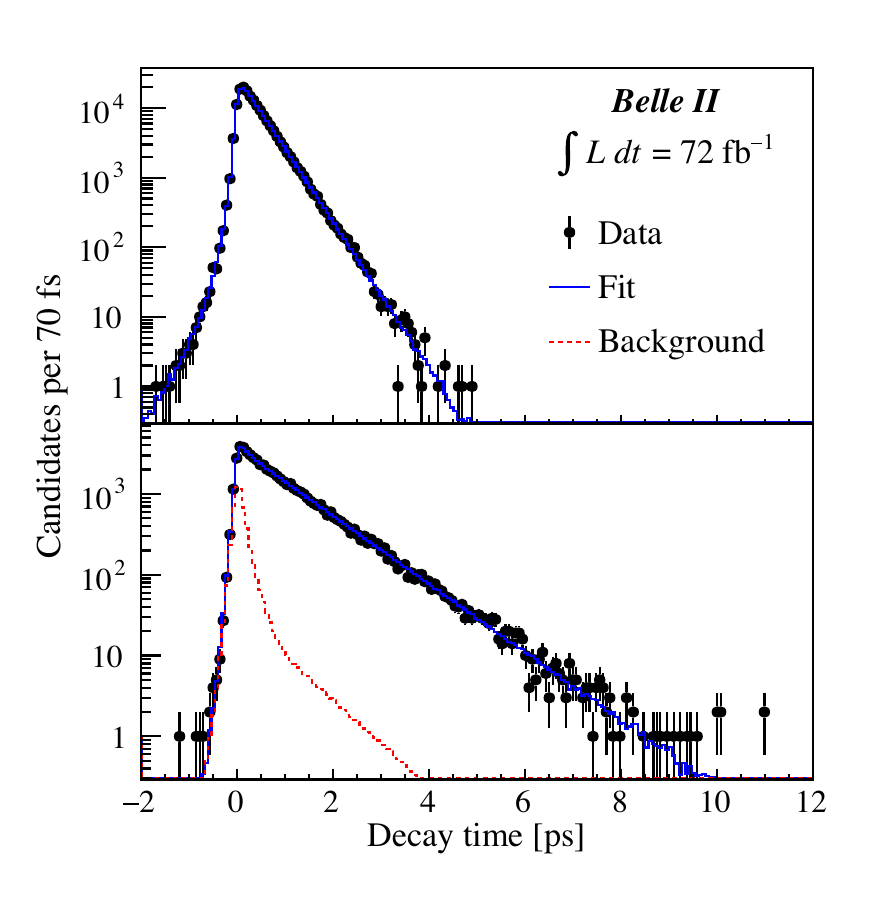}
\caption{\label{dlife} Decay-time distributions for $D^0\rightarrow K^-\pi^+$ (top) and 
$D^+\rightarrow K^-\pi^+\pi^+$ (bottom) candidates in the respective signal regions with fit projections 
overlaid~\cite{D}.}
\end{figure}

Using a data sample corresponding to 72 fb$^{-1}$, the lifetime of the $D^0$ and $D^+$ are measured to be
$410.5\pm1.1\pm0.8$ fs and $1030.4\pm4.7\pm3.1$ fs, respectively, where the first
uncertainty is statistical and the second is systematic~\cite{D}. The decay-time distributions with lifetime fit projections
overlaid are shown in Fig.~\ref{dlife}. These measurements are consistent with the
current world-averages, but with better precision. The sub-percent accuracy establishes the excellent
detector performance at Belle~II and paves the way for additional lifetime measurements.

The lifetime of the $\Lambda_c^+$ is measured with a data sample corresponding to 207.2 fb$^{-1}$ using
a relatively clean sample of $\Lambda_c^+\rightarrow pK^-\pi^+$ events. This data sample includes additional
data and an improved alignment calibration relative to the sample used for the $D^0$ and $D^+$ lifetimes.
Since the decay-time measurement assumes prompt production of $\Lambda_c^+$ candidates at the
interaction region of the $e^+e^-$ beams, backgrounds due to $\Xi_c^0\rightarrow\Lambda_c^+\pi^-$ and
$\Xi_c^+\rightarrow\Lambda_c^+\pi^0$ have the potential to bias the lifetime measurement. These backgrounds
were not accounted in the previous measurement by CLEO~\cite{Mahm} and are not relevant to the measurement
made at LHCb~\cite{Aaij}, which used $\Lambda_c^+$ candidates from semileptonic $\Lambda_b^0$ 
decays. While the branching fraction for $\Xi_c^0\rightarrow\Lambda_c^+\pi^-$ has been measured to be
$0.55\pm0.20$\%~\cite{Xic0}, no measurement has been made for the branching fraction for 
$\Xi_c^+\rightarrow\Lambda_c^+\pi^0$, though it was predicted to be about twice that of 
$\Xi_c^0\rightarrow\Lambda_c^+\pi^-$~\cite{Xicp}. The lifetimes of the $\Xi_c^0$ and $\Xi_c^+$ are $153\pm6$ fs
and $456\pm5$ fs, respectively~\cite{PDG}. To reduce backgrounds
from $\Xi_c$ decays, events with an invariant mass difference between the $\Xi_c$ and $\Lambda_c^+$ 
candidates within 2$\sigma$ of the expected value are rejected. A conservative estimate of the remaining
backgrounds is made by fitting the impact parameter of the $\Lambda_c^+$ candidate in the plane transverse
to the beam direction with a signal and background shape. Since the background shape includes both the
$\Xi_c$ and combinatorial backgrounds, it provides an estimated maximum contamination from $\Xi_c$
decays. Simulated $\Xi_c^0\rightarrow\Lambda_c^+\pi^-$ and $\Xi_c^+\rightarrow\Lambda_c^+\pi^0$ events
are then added to a simulated sample corresponding to five times the experimental data sample. The difference
in the $\Lambda_c^+$ lifetime extracted from a fit to the simulated sample with and without the $\Xi_c$ events
is 0.68 fs. Half of this shift is subtracted from the $\Lambda_c^+$ lifetime measured in the experimental data
as a correction. The full amount of the correction is taken as a systematic uncertainty associated with $\Xi_c$
backgrounds.

After the correction, the $\Lambda_c^+$ lifetime is determined to be $203.20\pm0.89\pm0.77$ fs, where the
first uncertainty is statistical and the second is systematic~\cite{Lc}. This measurement is consistent with the current
world average, but with better precision than previous experiments. The slight tension between the $\Lambda_c^+$
lifetime measurement by CLEO and other measurements remains. The Belle~II measurement provides a
benchmark for future baryon lifetime measurements.

\begin{figure}
\includegraphics[width=\linewidth]{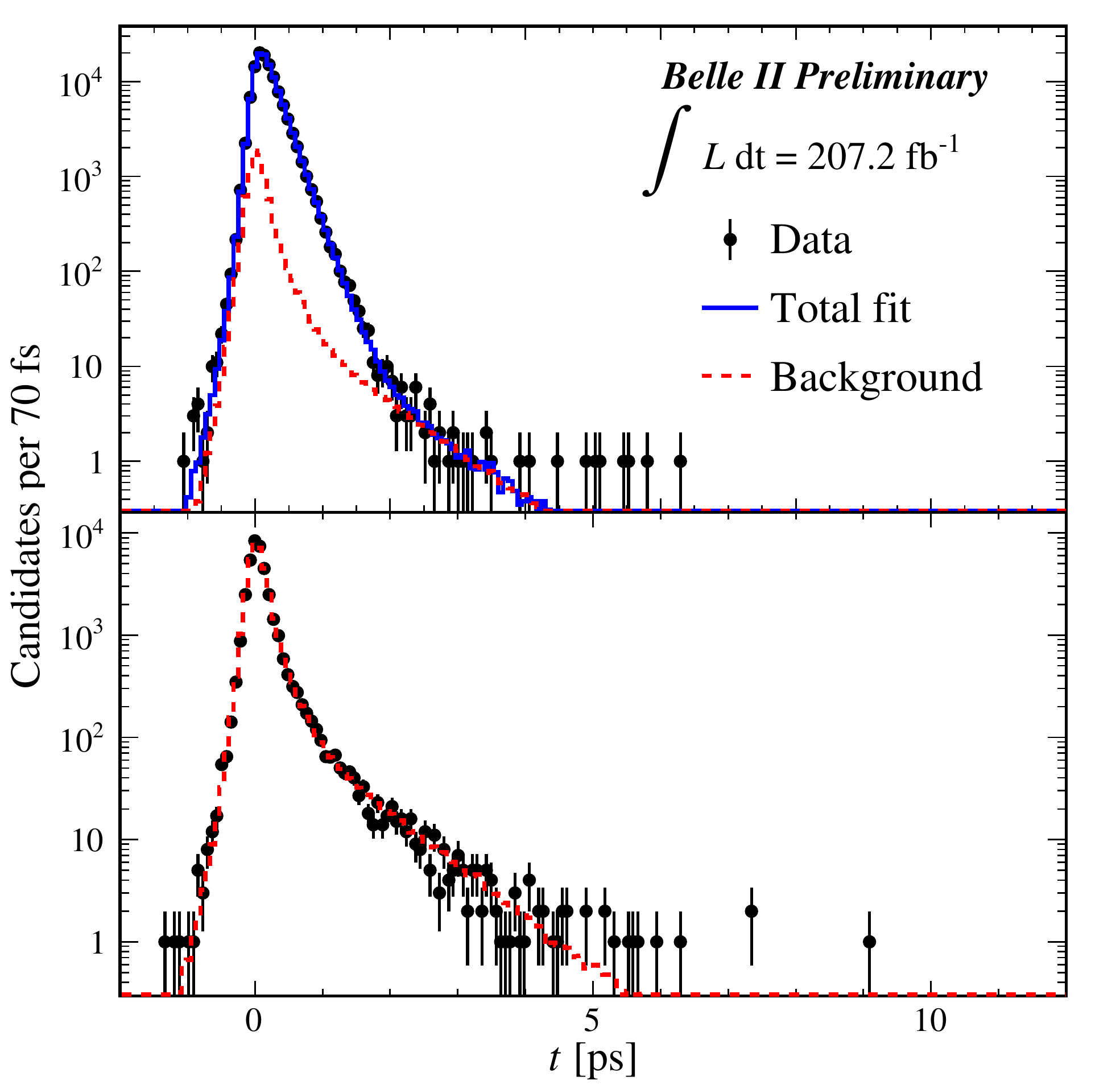}
\caption{\label{lclife} Decay-time distributions for $\Lambda_c^+\rightarrow pK^-\pi^+$ candidates in the signal 
region with fit projections overlaid~\cite{Lc}.}
\end{figure}

\section{Conclusions}
The Belle~II detector at the SuperKEKB collider benefit from upgrades in many detector and accelerator
components, in addition to improvements in electronics, software, and physics analysis tools, and mark the
beginning of the next-generation super-B-factory. Even with early data, Belle~II has shown the capability to
produce world-leading, high-precision results, including the world's best measurements of the lifetimes of the
$D^+$, $D^0$, and $\Lambda_c^+$. These measurements establish the excellent vertexing capabilities
of Belle~II and provide benchmarks for future lifetime measurements.

\bigskip 

\end{document}